\documentclass[12pt]{article}

\usepackage{color}
\usepackage{hyperref}  
\usepackage{graphicx}
\usepackage{amsfonts}
\usepackage{footmisc}
\usepackage{bm}

\usepackage{fancyhdr}
\let\a=\alpha \let\b=\beta  \let\g=\gamma  \let\d=\delta \let\e=\varepsilon
\let\z=\zeta  \let\h=\eta    \let\k=\kappa \let\l=\lambda
    \let\n=\nu    \let\x=\xi     \let\p=\pi    \let\r=\rho
\let\s=\sigma    \let\f=\varphi 
     \let\o=\omega
 \let\D=\Delta  \let\L=\Lambda \let\X=\Xi
     \let\F=\Phi    
\let\O=\Omega 

\font\tenmib=cmmib10\font\sevenmib=cmmib7\font\fivemib=cmmib5%
\mathchardef\Bl   = "0515  

\def\tto{\Rightarrow}
\def\Ba   {{\mbox{\boldmath$ \alpha$}}}

\def\Bd   {{\mbox{\boldmath$ \delta$}}}

\def\Bl   {{\mbox{\boldmath$ \lambda$}}}

\def\Bn   {{\mbox{\boldmath$ \nu$}}}

\def\Bs   {{\mbox{\boldmath$ \sigma$}}}

\def\Bff  {{\mbox{\boldmath$ \varphi$}}}

\def\Bo   {{\mbox{\boldmath$ \omega$}}}

\def\BDpr {{\mbox{\boldmath$ \partial$}}}

\def\eqalign#1{\null\,\vcenter{\openup\jot
  \ialign{\strut\hfil$\displaystyle{##}$&$\displaystyle{{}##}$\hfil
      \crcr#1\crcr}}\,}

\def\AA{{\mathcal A}}\def\CC{{\mathcal C}}
\def\DD{{\mathcal D}}\def\TT{{\mathcal T}}

\def\RR{\cal R}

\def\xx{{\V x}}

\def\hh{{\V h}}\def\qq{{\V q}}\def\pp{{\V p}}
\def\T#1{{#1_{\kern-3pt\lower7pt\hbox{$\widetilde{}$}}\kern3pt}}

\def\ie{{\it i.e.\ }}
\def\dpr{{\partial}}
\def\defi{{\buildrel def\over=}}

\def\otto{\,{\kern-1.truept\leftarrow\kern-5.truept\to\kern-1.truept}\,}
\def\Pprod{\prod^{\kern-1mm\raise.0mm\hbox{$\leftarrow$}}}
  
\newdimen\xshift \newdimen\xwidth \newdimen\yshift \newdimen\ywidth

\def\ins#1#2#3{\vbox to0pt{\kern-#2pt\hbox{\kern#1pt #3}\vss}\nointerlineskip}

\def\eqfig#1#2#3#4#5{
\par\xwidth=#1pt \xshift=\hsize \advance\xshift
by-\xwidth \divide\xshift by 2
\yshift=#2pt \divide\yshift by 2
{\hglue\xshift \vbox to #2pt{\vfil
#3 \includegraphics{#4.eps}
}\hfill\raise\yshift\hbox{#5}}}

\def\V#1{{\bf #1}}
\def\lis#1{{\overline#1}}
\let\wt=\widetilde
\let\wh=\widehat

\def\tende#1{\,\vtop{\ialign{##\crcr\rightarrowfill\crcr
 \noalign{\kern-1pt\nointerlineskip} \hskip3.pt${\scriptstyle
   #1}$\hskip3.pt\crcr}}\,}
\def\eg{{\it e.g.\ }}

\def\0{\noindent}
\def\*{\vskip2mm}

\def\Eq#1{\label{#1}}
\def\equ#1{(\ref{#1})}

\def\iniz{\setcounter{equation}{0}}
\def\be{\begin{equation}}\def\ee{\end{equation}}
\renewcommand{\theequation}{\arabic{section}.\arabic{equation}}

\def\be{\begin{equation}}\def\ee{\end{equation}}
\renewcommand{\theequation}{\arabic{section}.\arabic{equation}}

\newcounter{appendice}
\def\APPENDICE#1{
\setcounter{appendice}{#1}
\appendix
\renewcommand{\theequation}{\alph{appendice}.\arabic{equation}}%
\renewcommand{\thesection}{\Alph{appendice}}%
}

\def\alert#1{{\color{ired}#1}}
\definecolor{iblue}{RGB}{65,105,225}
\definecolor{ired}{RGB}{220,20,60}
\definecolor{igreen}{RGB}{50,205,50}
\definecolor{ipurple}{RGB}{75,0,130}
\definecolor{iochre}{RGB}{218,165,32}
\definecolor{iteal}{RGB}{51,204,204} 
\definecolor{imauve}{RGB}{204,51,153}

\def\tto{\Rightarrow}

\def\pp{{\bf p}}\def\xx{{\bf x}}\def\rr{{\bf r}}\def\hh{{\bf h}}
\def\eee{{\bf e}}\def\KK{{\cal K}}\def\nn{{\bf n}}
\def\aa{{\bf a}}\def\AA{{\V A}}\def\zz{{\bf z}}\def\qq{{\bf q}}

\def\ZZ{\mathcal Z}
\begin{document}

\title{\alert{\bf Ergodicity, KAM, FPUT}}                 

\author{Giovanni Gallavotti                
  \\ {\it Dipartimento di Fisica,}         
  \\ {\it Universit\`a di Roma-La Sapienza and  I.N.F.N.-Roma1}
  }                                        
\date{}                                    
\maketitle
\begin{abstract}
  Boltzmann introduced the microcanonical ensemble in
  1868, \cite{Bo868-a}, and immediately attempted to give an
  example of a system whose stationary states would be described
  by the emsemble (as suggested also by his ergodic
  hypothesis). The example, \cite{Bo868-b}, has been recently
  shown to be incorrect, if taken literally: the point was to
  suppose that constants of motion, if any besides the energy,
  would necessarily be smooth functions; and soon later he warned
  on the dangers implicit in a similar assumption. Fifty years
  later Fermi wrote a paper attempting to prove that in general a
  nonlinear system should be ergodic, \cite{Fe923}: but his proof
  relied again on Boltzmann's assumption.  Thirty-four more years
  elapsed, and Fermi returned on the problem collaborating with
  Pasta, Ulam, Tsingou: the surprise was that the considered non
  linear chain was apparently not following the ergodic
  hypothesis. In the same year Kolmogorov had proved the
  conservation of many quasi periodic motions in nonlinear
  perturbations of integrable systems, \cite{Ko954}: his theorem was
  considered, already a few years later, a possible explanation
  of the FPUT work, \cite{Fo992}. This was only the beginning of intense
  research: here a brief sketch is presented to illustrate the
  above themes and the connection with the multiscale aspects of
  the problems, and the ``Renormalization group method'' intended
  as a map $\RR$ whose iterations can be interpreted as
  successive magnifications, zooming on ever smaller regions of
  phase space in which motions develop closer and closer to the
  searched quasi periodic motion of given spectrum.
  
\end{abstract}

\setcounter{section}{0}
\def\SEC{Ergodicity}
\section{\SEC}                             
\label{sec1}\iniz                          
\lhead{\SEC}     

The Hamiltonian of the systems of $N$ oscillators $r_x = (p_x,r_x),
x = 1, ..., N$ , considered by FPUT was, ($N = 64, r_0 = r_N =
0$):
\be\sum^{N-1}_{x=1}
\frac{p_x^2}{2m}+
\sum_{x=1}^{N-1}
\frac12(r_{x+1}-r_x)^2
+\frac\a3(r_{x+1}-r_x)^3
+\frac\b4(r_{x+1}-r_x)^4
\Eq{e1.1}\ee
representing a  linear chain of oscillators, which can be
generalized to a $d$-dimensional lattice with
$r,x$ as $d$-dimensional vectors, for instance:

\be \sum_{\xx\in Z_N}\frac12{\pp^2_\xx}
+\sum_{\xx\in Z_N}\sum_{i=1}^d
\frac12(\rr(\xx+\eee_i)-\rr_\xx)^2+\frac\b2
(\rr_(\xx+\eee_i)-\rr_\xx)^4
\Eq{e1.2}
\ee
where $\eee_i,i=1,\ldots,d$ is an ortonormal basis and
$Z_N=(0,\ldots,N-1)^d$ is the cube with side $N$ in the lattice
$Z^d$, and $\rr_\xx=0$ if any coorinate of $\xx$ is $0$ or $N$.

Boltzmann's ergodic hypothesis, \cite{Bo866}, formally stated an
hypothesis, which had been perhaps earlier conceived (for
instance by Descartes, \cite[p.4]{Da014}) and which had been
accepted by Maxwell, \cite{Ma879}, supposes that the point
representing the system istantaneous state wonders visiting all
points that can possibly be visited (\ie compatible with the
constraints).

The hypothesis, leaving aside the obvious exceptions, and its
implication that all points covered in time a closed orbit
allowed Boltzmann to discover a mechanical interpretation of the
Second Law as a manfestation of the Maupertuis' {\it least
  action} principle via the following extraordinary argument.
\*

\0(i) An equilibrium state of a system is idetified with
a periodic motion $t\to x(t)$ developing under the action of
internal forces with potential energy $V(x(t))$ and of external
forces with potential $V_{ext}(x(t))$; its kinetic energy is
$K(x(t))$, and the {\it action} of $x$ is defined, if its period
is $i$ (adopting here the original notation for the period), by

\be{\cal A}(x)=\int_0^i \big(\frac{m}2\dot x(t)^2
-V(x(t))\big)\,dt\Eq{e1.3}\ee

\0(ii) Properties of a small periodic variations $\d x(t)$ of a state,
\ie properties of an equilibrium state close to $x$, represented  as:
\be\d x(t)=x'(\frac{i'}i t)-x(t)\defi
x'(i'\f)-x(i\f)\hfill\Eq{e1.4}\ee
where $\f\in[0,1]$ is the {\it phase}, as introduced by Clausius,
\cite{Cl871}.
The role of $\f$ is simply
to establish a correspondence between points on the initial trajectory $x$
and on the varied one $x'$: it is manifestly an arbitrary correspondence
(which could be defined differently without affecting the final result)
convenient to follow the algebraic steps.\footnote{\small It should be noted that
Boltzmann does not care to introduce the phase and this makes his
computations difficult to follow (in the sense that once realized which
the final result is, it may be easier to reproduce it rather than to follow his
calculations).}
\*

\0(iii) Following Clausius version, \cite{Cl871}, the heat theorem is
deduced from a few identities about the average values
$\lis{F}=i^{-1}\int_0^i F(x(t))dt$ for generic observables $F$ and
$\lis K= i^{-1}\int_0^i\frac12\dot x(t)^2dt$ for the average of the
kinetic energy.

Let {$x(t)=\x(\frac{t}i)$ with $\x(\f)$ 1-periodic}; it is $\lis
F\equiv \int_0^1 F(\x(\f))d\f$, and $\lis K=\frac12\int_0^1
\frac{\x'(\f)^2}{i^2} d\f$.  Changing $x$ by $\d x$ means to
change $\x$ by $\d\x$ and $i$ by $\d i$. So variation of an
average is the sum of the variation due to $\d\x$ and that due to
$\d i$: namely {$\d \lis F=\d_\x {\lis F}+\d_i \lis
  F$}. Therefore
\be{\d\lis V\equiv\d_\x{\lis
    V}},\qquad \d\lis{V_{ext}}=\d_\x{\lis{V_{ext}}},\qquad {\d
  \lis K=-2\frac{\d i}i \lis K +\d_\x \lis K}\Eq{e1.5}\ee
The variation $\d (\lis K +\lis V+\lis{V_{ext}} )\defi dU+dL$, with
$U\defi\lis K +\lis V$ and $dL\defi d \lis{V_{ext}}$, has a natural
interpretation of heat $\d Q$ in/out in the process $x\to
x'$.\,\footnote{\small Because $\d (\lis K +\lis V)\defi dU$ and
$\d \lis{V_{ext}}\defi \d L$ are the variation of the internal
energy $U$ and $dL$ is the work that the system performs: \ie $\d
Q=\d U+\d L$.}

Then using the above Eq.\,\equ{e1.5}:
$\d_\x \lis K=\d \lis K+2\frac{\d i}i \lis K $
\be\eqalign{
&\d Q\equiv { \d (\lis K +\lis V+\lis{V_{ext}} )
=-2\frac{\d i}i\lis K+\d_\x ({\lis K}+\lis V+\lis{V_{ext}} )}
\cr
&
={\bf-}2\frac{\d i}i\lis K+{2\d_\x{\lis K}}+
\Bd_\x {\bf ({-{\lis
      K}} +\lis V+\lis{V_{ext}})}\equiv
   {\bf +}2\frac{\d i}i\lis K+2\d {\lis K} +{\bf 0}\cr}
\Eq{e1.6}\ee
where the last ${\bf 0}$ is due to {\it Maupertuis' principle},
as the motion follows the equations of motion (at fixed $i$).
Therefore, ({\it Heat theorem}):
\be{\frac{\d Q}{\lis K}}=+2\frac{\d i}{i}+2\frac{\d\lis K}{\lis K}=
{2\d\log (i\lis K)}\defi{\bf \d S}\Eq{e1.7}\ee
is an {exact} differential.
In reality it is somewhat strange that both Boltzmann and
Clausius call the last equation a ``generalization of the action
principle'': the latter principle uniquely determines a motion,
{\it i.e.} it determines its equations. The equation instead
does not determine a motion but it only establishes a relation
between the variation of average kinetic and potential energies
of close periodic motions under the assumption that they satisfy
the equations of motion; and it does not establish a variational
property, see \cite[Sec.I.4]{Ga025}.

The above analysis was successive to the ergodic hypothesis and a
complement to it: after 
achieving the construction of the microcanonical
ensemble, \cite{Bo868-a}, as recognized by Maxwell in one of his
last works,\cite{Ma879}, and later by Gibbs, \cite{Gi902} (who
refers to a later publication, \cite{Bo871-b}, curiously quoting
it under the title of the first section) and soon Boltazmann
wanted to provide simple exapmles of ergodic systems.

He examined the system of one particle, in dimension $2$, subject
to a central force of potential $-\frac{\a}{|\xx|}$, to a
potential $+\frac{\b}{|\xx|^2}$ and to a flat repelling line.
For $\b=0$ the motion is integrable and Boltzmann tried to prove
that as soon as $\b>0$ the system would be ergodic. This
is not right (as conjectured in \cite{Ga016,GJ020} and proved in
\cite{Fe021}) the example fails as a consequence  of
the KAM theorem, discovered in the early '950s by Kolmogorov,
\cite{Ko954}. Nevertheless Boltzmann's paper was probably known to Fermi,
who in his early youth attempted to provide a proof of the
ergodic hypothesis,\cite{Ga004c} (overlooking a similar difficulty: \ie the
possibility of the existence of non smooth constants of motion);
hence when (relatively) large computers became available the
FPUT project was attempted.

Existence of non smooth constants of motion rests, ultimately,
upon realizing that the phase space may have a large volume
covered by smooth invariant surfaces of low dimension: however
proving their existence requires a multi scale analysis: by
examining the motions with ever increasing precision, invariant
curves can be separated from the nearby motions (which are likely
to undergo chaotic motions).

\def\SEC{KAM}
\section{\SEC}                             
\label{sec2}\iniz                          

In spite of its interest the first attempt to establish a
rigorous connection between the problem of existence of non
smooth constant of motions and the ergodicity of the FPUT chain
had to wait 1971, \cite{Ni971}, treating ``only'' $N$ particles
chains with $N$ either prime or a power of $2$ (hence including at least
the cases studied in FPUT work $N=16,32,64$!) and a
complete proof, including the ``other $N$-s'' ($N$ not a power of
$2$ or not a prime), dates 2006, \cite{Ri006}; both, although
conceptually very different, relied on the
KAM theorem, which is discussed below as an application of the
renormalization group method.

Consider the Hamiltonian:
\be H_0(\V A,\Ba)=\frac12 (\V A\cdot J_0\V A)+\Bo_0\cdot{\V A}
+f_0(\V A,\Ba)\equiv h_0+f_0\Eq{e2.1}\ee
real analytic for $(\V A,\Ba)\in (\DD_\r\times \TT^N)$, $N\ge2$, with:
$\DD_\r=\{\V A\in R^N, |A_j|<\r\}$, $\TT^N$ =
$N$-dimensional torus $[0,2\p]^N$, identified with unit
polydisk, $\{\V z |\, z_j=e^{i\a_j}, j=1,\ldots,N\}$, $\Bo_0\in
R^N$ and $J_0$ is a $N\times N$ matrix, positive for simplicity,
with eigenvalues in $[J_{0,-},J_{0,+}]$.
                                           
The Hamiltonian is supposed holomorphic in the complex region
$\CC_{\r_0,\k_0}=\{(\AA,\zz)| |A_j|\le\r_0, e^{-\k_0}\le|z_j|\le
e^{\k_0}\}$; the label $0$ is attached, since the
beginning, because $H_n,f_n,\r_n,\k_n$ will arise later, with
$n=1,2,\ldots$; the perturbation $f_0$ size will be measured by
$\e_0$ as defined here:                                   
\be\eqalign{                               
  \CC_{\r_0,\k_0}&\defi\{(\V A,\V          
  z)||A_j|\,\le\,\r_0,\ e^{-\k_0}\le| e^{i\a_j}|\le
  e^{\k_0}, j=1,\ldots,N\}\subset \CC^{2N}\cr
  \e_0&=||\BDpr_{\V A}f_0||_{\r_0,\k_0}   
  + \frac1{\r_0} ||\BDpr_{\Ba}f_0||_{\r_0,\k_0}
  \qquad{\rm with:}\cr                    
  ||f&||_{\r_0,\k_0}\defi\max_{\CC_{\r_0,\k_0}} |f(\V A,\V z)|,
  \qquad \forall f\ {\rm holomorphic\ in}\ \CC_{\r_0,\k_0} \cr}
\Eq{e2.2}\ee
with $\r_0>0,1\ge\k_0>0$, $z_j\equiv e^{i\a_j}$; generally
$\CC_{\r,\k}(\V {\lis A})$ will denote a 'polydisk' centered at
$\V {\lis A}$, \ie defined as in Eq.\equ{e2.2} with $|A_j-\lis
A_j|\le \r$ replacing $|A_j|\le \r$ and $\k$ replacing $\k_0$;
polydisks centered at the ``origin'' $\lis\AA=0$ will be simply denoted
$\CC_{\r,\k}$ and called ``centered polydisks'', while $C_\r(\lis \AA)$ is
the $\{\AA| |A_i-\lis A_i|<\r\}$.  Set $|\V
A|=\max |A_j|, |\V z|=\max|z_j|,\ \forall \V A,\V z\in \CC^N$.
\*

{\it For simplicity functions of $\Ba$ will be implicitly
  regarded as functions of $z_j=e^{i\a_j}$ and their arguments
  will be written as $\V z$ or $\Ba$, as convenient.}  \*

The idea is to focus attention on the center of $\CC_{\r_0,\k_0}$
where, if $\e_0=0$, a motion (``free motion'': $(\V A,\Ba)\to (\V
A,\Ba+\Bo_0 t)$) takes place which is quasi periodic ``with
spectrum'' $\Bo_0$. This is done by studying motions in a small
polydisk $\CC_{\wt\r,\wt\k}(\aa)\subset \CC_{\r_0,\k_0}$, eccentric
if $\aa\ne\V 0$.

Thus motions starting in
$\CC(\wt\r,\wt\k)(\aa)\subset\CC(\r_0,\k_0)$ near the center of
$\CC(\r_0,\k_0)$ can be studied as ``through a microscope'': in
the good cases (\ie under suitable assumption on the initial
parameters $J_0,\Bo_0$ and $f_0$) the Hamiltonian will turn out
to be expressible (after a further coordinate change, to turn
$\CC(\wt\r,\wt\k)(\aa)$ into $\CC(\r_1,\k_1)$, centered at $0$)
in a form substantially
closer to that of a quasi periodic oscillator (described by its
``normal'' Hamiltonian $\Bo_0\cdot \V A$, or $\Bo_0\cdot \V
A+\frac12 \AA\cdot J_0\AA$, in the variables $\V
A,\Ba$).

Iterating the process the Hamiltonian changes but, {\it remaining
  analytic in a polydisk $C_{\r_n,\k_n}$}, converges to that of a
  harmonic oscillator: the interpretation will be that, looking
  very carefully in the vicinity of the ``unperturbed'' torus
  $\TT_{\Bo_0}=\{\V A=\V 0,\Ba\in[0,2\p]^N\}$, the perturbed
  Hamiltonian exhibits with increasing precision a harmonic
  motion with spectrum $\Bo_0$.

The result is the KAM theorem in a form not only
reminiscent of the methods called ``renormalization group'', RG,
in quantum field theory but just a realization of them,
\cite{Ga986}. Existence of the invariant torus can be seen as the
existence of a trivial fixed point (a harmonic oscillator) of a
single canonical coordinate change together with a time
rescaling.
\*

The ``spectrum'' $\Bo_0$ will be supposed
``diophantine'', \ie for some $C>0$ it is, denoting $\ZZ^N$
the lattice of the integers, for all $\V 0\ne\Bn\in \ZZ^N$:
\be |\Bo_0\cdot\Bn|\, \ge \,C\, |\Bn|^{-N},
\qquad  |\Bn|\equiv\sum_{i=1}^N |\n_i|> 0\Eq{e2.3}\ee
The inequality Eq.\ref{e2.3} will be repeatedly used to define
canonical transformations $(\AA,\Ba)\otto(\AA',\Ba')$ with
generating functions $\F(\V A',\Ba)+(\V A'+\aa)\cdot\Ba$:
\be \V A=\V A'+\aa+\BDpr_\Ba \F(\V A',\Ba), \quad
\Ba'=\Ba+\BDpr_{\V A'}\F(\V A',\Ba)\Eq{e2.4}\ee
with $\F$ chosen so that in the new coordinates $(\V A',\Ba')$
the perturbation is '{\it weaker}'; at the price that the new
coordinates will cover a (much) smaller domain, inside
$\CC(\r_0,\k_0)$.

The $\aa$ is introduced because the quasi
periodic motion with spectrum $\Bo_0$ which, in the first
approximation with $f_0=0$, is located at $\AA=0$, at a better
approximation is shifted, as expected at least if the
$\Ba$-average $\lis f_0(\AA)$ of $f_0(\AA,\Ba)$ is not
$0$:\,\footnote{\small As shown already by the cases with $f_0$ not
depending on $\Ba$.}\, the constant $\aa$ provides flexibility
to identify the average position, in $\V A'$-space, of the new
approximations.

Suppose that a holomorphic $\F(\AA',\Ba)$ defines, via
Eq.\equ{e2.4}, in a domain $(\AA',\Ba')\in\CC_{\wt\r,\wt\k}$, a
symplectic map $(\AA',\Ba')\to (\V A,\Ba)$ with image
$\subset\CC_{\r_0,\k_0}$.

{\it Choose} $\aa$ as solution of the implicit equation
$\aa=-J_0^{-1}\BDpr_\aa \lis f_0(\aa)$ and insert Eq.\equ{e2.4}
into $H_0$.

After easy\&patient algebra, the Hamiltonian becomes,
\,\footnote{\small\label{imagine} $\Ba$ has to
be imagined as a function of $(\AA',\Ba')$, solution of the
second implicit equation in Eq.\ref{e2.4}.}\, up to a constant,
the sum of the four lines (that will be referred as $F_j,
j=0,\ldots,4$):
\be\eqalign{(0):\quad&H'(\V A',\Ba')=\frac12 (\V A'\cdot J_0\V
  A')+\Bo_0\cdot\V A'\cr
  (1):\quad&+\lis f_0(\V A'+\aa)-\lis f_0(\aa)-\BDpr_{\aa}\lis f_0(\V
    a)\cdot\V A'\cr
  (2):\quad&+ \Big(f_0(\V A'+\aa+\BDpr_\Ba\F,\Ba)-f_0(\V A'+\V
    a,\Ba)\Big)+\frac12\BDpr_\Ba\F\cdot J_0\BDpr_\Ba\F\cr
  (3):\quad&+(\Bo_0+J_0(\V A'+\aa))\cdot \BDpr_{\Ba}\F
    +f_0(\V A'+\aa,\Ba)-\lis f_{\V 0}(\V A'+\aa)\cr
    }\Eq{e2.5}\ee
where a few terms have been added or subtracted (including free
addition or subtraction of constants, like $\Bo_0\cdot \aa$,
$\frac12J_o\aa\cdot\aa$ and $\lis f_0(\aa)$): so that each line
$F_j,j>0$, can {\it formally} be estimated to have size $\sim
O(\e_0^2)$ in a smaller polydisk, while the line $(0)$ is the new
unperturbed Hamiltonian (in the $(\AA',\Ba')$
variables\,\footref{imagine}\,).\,\footnote{\small
Heuristically: line (1) is a sum of quantities of order $\e_0$
but it is of second order in $\V A'$ and could be made of
$O(\e_0^2)$ by suitably restricting size of $\V A'$: remark that
$\lis f_0(\aa)$ is a constant added freely but the last term is
there because of the choice $\aa=-J_0^{-1}\BDpr_{\aa}\lis
f_0(\aa)$.  Line (2) can be made of $O(\e_0^2)$ by choosing $\F$
of $O(\e_0)$. Line (3) could even be set $=0$ and invites taking
it a {\it definition} of $\F$ (if possible).}

As familiar in pertubation theory this may suggest to
define:
\be \eqalign{
 &\aa=-J_0^{-1}\BDpr_{\aa}\lis f_0(\aa)\cr
 & \F(\V A',\Ba)=-\sum_{\V 0\ne\Bn\in \ZZ^N}
  \frac{f_{0,\Bn}(\V A'+\aa)}{ i(\Bo_0\cdot\Bn+((\V A'+\V
    a)\cdot J_0 \Bn))} e^{i\Bn\cdot\Ba},\cr
    }
  \Eq{e2.6}\ee
Existence of $\aa$ inside $C_{\r_0}$ can be easily established
via {\it dimensional bounds}.

\begin{itemize}
\item In general an example of dimensional bound is a bound of a
holomorphic function $f(\z)$ and of its derivatives at a point
$\z$ in terms of its maximum $|f|$ in the definition domain and
of the distance $d(\z)$ of $\z$ to the boundary: the
$n^{th}$-derivative at $\z$ is bounded by $n!(\max |f|)\,
d(\z)^{-n}$.  Likewise a function $g(\z)$ holomorphic
in an annulus bounded by the circles
$\z=e^{i\a\pm\k},\a\in[0,2\p]$, will have Fourier's coefficients
bounded by its maximum $\e$ in the annulus as $|g_{\Bn}|\le \e
\exp(-\k\,|\Bn|)$. The mentioned bounds will be called 
dimensional bounds.

\item In conclusion, if $\e=\max_{\CC(\r,\k)} |\BDpr_\AA
F(\AA,\Ba)|+\max_{\CC(\r,\k)}\frac1\r |\BDpr_\Ba F(\AA,\Ba)|$ and $F$ is
holomorphic in $\CC(\r,\k)$, a dimensional bound on $F(\AA,\Ba)$
for $(\AA,\Ba)\in\CC((1-\l)\r,\k)$, $\l<1$, is, for $\Bn\ne0$:
\be |\BDpr_\AA^\nn \BDpr_\AA F(\AA,\Ba)|< \nn!\,
(\l\r)^{-|\nn|}\,\e ,\qquad |\Bn|
| F_\Bn
(\AA,\Ba)|<e^{- \k\, |\Bn|}\, \r\, \e \Eq{e2.7}\ee
where $\nn\in\ZZ_+^N,\,|\nn|=\sum n_i,\,\nn!=\prod
n_i!,\,\Bn\in\ZZ^N,\,|\Bn|=\sum_i |\n_i|$.\,\footnote{\small
Remark that to estimate $\AA$-derivatives the $\r$ has to be
reduced by a factor $1-\l$ while to estimate $\Ba$-derivatives
$\k_0$ can be left unchanged, as such $m$-th $\Ba$-derivatives
modify the second of Eq.\ref{e2.7} by a factor $|\Bn|^m$.}

\item In the following, dimensional bounds will be
used repeatedly: when a bound on a function holomorphic in
$\CC(\r,\k)$ is needed, it will be obtained dimensionally but
restricted to a smaller polydisk $\CC(\l\r,\k-\d)$, for
$\l<1,\d<\k$ (conveniently fixed each time, {\it without} attention to
optimize the choice to obtain better bounds).
\*

\item As a first example, a condition for the existence of the shift
$\aa$, in the first of Eq.\equ{e2.6}, is $\e_0$ small:
{\it if $J_{0,-}^{-1}\e_0 \r_0^{-1}<\frac1{\g_0}$ and $\g_0$ is
large enough, the equation $\aa=-J_0^{-1}\BDpr_\aa\lis f(\aa)$
has a solution $\aa$ with $|\aa|< J_{0,-}^{-1}\e_0<\frac1{\g_0}\r_0$}: see
Appendix\,\ref{A}\, where the choice $\g_0=3N$ is proposed.
\end{itemize}

\*

However the $\F$ in Eq.\equ{e2.6} does not make sense (in
 general); {\it instead} it will ``be approximated'', and {\it
 properly defined}, by $\F_0(\V A',\Ba)$ as:
\be\kern-3mm\eqalign{&-\kern-8pt\sum_{\V 0\ne\Bn\in \ZZ^N}
  \kern-5pt\frac{f_{0,\Bn}(\V A'+\aa)
   e^{i\Ba\cdot\Bn}}{i\Bo_0\cdot\Bn} {\Big(1-\frac{J_0(\V A'+\V
   a)\cdot\Bn}{\Bo_0\cdot\Bn}+ \Big(\frac{J_0(\V A'+\V
   a)\cdot\Bn}{\Bo_0\cdot\Bn}\Big)^2\Big)}\cr}\Eq{e2.8}\ee
\0From
   $\e_0=\|\dpr_{\AA'}f_0\|_{\r_0,\k_0}+\frac1{\r_0}\|\dpr_{\Ba}f_0\|_{\r_0,\k_0}$
   the dimensional inequalities (Eq.\ref{e2.7}):
\be|\BDpr_{\AA'}f_{0,\Bn}(\AA'+\aa)|\le \e_0
   e^{-\k_0|\Bn|},\qquad \r_0 |\Bn| |f_{0,\Bn}(\AA'+\aa)|<\e_0
   e^{-\k_0\cdot|\Bn|}\Eq{e2.9}\ee
hold if $\AA'+\aa\in C(\r_0)$; and allow
   to estimate $\F_0$ and to define properly the
   canonical map in Eq.\ref{e2.4} as a map from
   $\CC(\frac14\r_0,\k_0-2\d_0)$ to
   $\CC(\r_0-\frac12\r_0,\k_0-\d_0)$ for some analyticity loss
   parameter $0<\d_0<\frac14\k_0$.
\*

This is done via a few dimensional bounds obtained by successive
   reduction of analiticity domains $\CC(\r,\k)$ to
   $\CC(\r',\k')$ conveniently chosen.\,\footnote{\small Warning:
   every time a new estimate is proposed, the dimensionless
   constants involved in it will typically be $\g_j,c_j$ with a
   new label $j$ even when equal to constants with label $j'<j$.}

If $\g_0=3N$, as above, and $C$ is the
   Diophantine constant, Eq.\equ{e2.3}, the
\be J_{0,-}^{-1}\e_0< \frac1{\g_0}\r_0,\qquad
   J_{0,+}\r_0C^{-1}<1\Eq{e2.10}\ee
and the second of Eq.\equ{e2.9} imply a simple bound on
$\|\F_0\|_{\frac23\r_0,\k_0}$ \footnote{\small As $\AA'\in
C(\frac23\r_0)$, implies $\AA'+\aa\in C(\frac56\r_0)$: remark
$|\aa|<\frac1{3N}\r_0<\frac16\r_0$ for $N\ge2$.}\,.  Therefore
\be\kern-3mm\eqalign{
&\|\F_0\|_{\frac23\r_0,\k_0}\le
\e_0\r_0\kern-2mm \sum_{\Bn\ne\V0}
\frac{|\Bn|e^{-\k_0|\Bn|}}{|\Bo_0\cdot\Bn|}\Big(1\kern-2mm+
\frac{|J_0\Bn|\r_0}{|\Bo_0\cdot\Bn|}+\kern-2mm
\Big(\frac{|J_0\Bn|\r_0}{|\Bo_0\cdot\Bn|}\Big)^2\Big)
\le \g_1 \frac{\e_0}C \,\k_0^{-c_1}\cr
}\Eq{e2.11}\ee
which yields
\be\eqalign{
&||\BDpr_{\AA'}\F_0||_{\frac12\r_0,\k_0}+\frac1{\r_0}
\|\BDpr_\Ba\F_0\|_{\frac23\r_0,\k_0}\cr
&\le \wt\g\e_0 \sum_{\Bn\ne\V0}
\frac{|\Bn|^2e^{-\k_0|\Bn|}}{|\Bo_0\cdot\Bn|}\Big(1+
\frac{|J_0\Bn|\r_0}{|\Bo_0\cdot\Bn|}
+\Big(\frac{|J_0\Bn|\r_0}{|\Bo_0\cdot\Bn|}\Big)^2\Big)
\le \wh\g \frac{\e_0}C \,\k_0^{-\wh c}\cr}\Eq{e2.12}\ee
with $\wt\g,\wh\g,\wh c$ dimensionless constants.\,\footnote{\small
E.g. $\wh c=4N+2$, $\wh g=\wt\g\sum_\Bn |\Bn|^{4N+2} e^{-\k_0\,|\Bn|}$,
$\wt\g=7$. Remark that the bound on the second term in the first
line of Eq.\ref{e2.12} does not involve a change in $\k_0$, unlike the
first term (where $\r$ is reduced).}
\*

The bounds Eq.\equ{e2.12} lead easily, via 
dimensional estimates, to a precise definition of the canonical
map in Eq.\ref{e2.4} as a map $(\AA',\Ba')\in \CC(\wt\r,\wt\k)\to
(\AA,\Ba)\in \CC(\r_0,\k_0)$, where $\wt\r<\r_0,\wt\k<\k_0$ will
be suitably defined. And, once the canonical map is defined, the
Hamiltonian will take the form $H_1(\AA',\Ba')=
\Bo_0\cdot\AA'+\frac12\AA'\cdot J_0\AA'+ f_1(\AA',\Ba')$ for
$(\AA',\Ba')$ in the definition domain $\CC(\wt\r,\wt\k)$.

The 'new' perturbation will be, by Eq.\equ{e2.5},
$f_1(\AA',\Ba')=\sum_{j=1}^3 F_j(\AA',\Ba)$ if expressed in the
{\it mixed variables} $(\AA',\Ba)$. Hence it will eventually be
necessary to express $\Ba$ via $(\AA',\Ba')$ by inverting the
second in Eq.\ref{e2.4} as $\Ba=\Ba'+\D(\AA',\Ba')$: so that the
perturbation acting in $H_1$ will be, properly,
$f_1(\AA',\Ba')=\sum_{j=1}^3 F_j(\AA',\Ba'+\D(\AA',\Ba'))$.

To solve for the implicit function
$\D(\AA',\Ba')$ in the second of Eq.\ref{e2.4}\,, an argument can
be used similar to the one for the solution of
$\aa=-J_0^{-1}\BDpr_\aa\lis f_0(\aa)$ in Appendix\,\ref{A}\,.  In
Appendix\,{B} the similar analysis is adopted to construct
$\D(\AA',\Ba')$ and a proof that it can be defined on
$\CC(\frac12\r_0,\k_0-2\d_0)$ subject to the bound:
\be\|\D\|_{\frac12\r_0,\k_0-2\d_0}<\lis\g\frac{\e_0}C \,\k_0^{-\lis
c}\Eq{e2.13}\ee
for dimensionless  $\lis\g,\lis c$, and for $\d_0>0$ to be
determined and depending on $\e_0,\k_0$.
\*
However it is convenient to postpone the study of $\D$ and to obtain
first bounds on $f'_1(\AA',\Ba)=\sum_{j=1}^3 F_j(\AA',\Ba)$ in a
domain $\CC(\wt\r,\k_0)$.
\*
The conditions for the definitions of $F_j(\AA',\Ba)$, \ie for
the definition of the generating function $\F_0$ to be defined in
the mixed variables $(\AA',\Ba)$, are summarized in
Eq.\ref{e2.10}

Under such conditions the $F_j$ composing
$f'_1(\AA',\Ba)$, in the lines labeled $(j),j>0$ in Eq.\equ{e2.5}, can be
easily bounded dimensionally as follows.
\*

\begin{itemize}
\item
\0(a) $F_0$ is $\frac12(J_0\AA'\cdot\AA')+\Bo_0\cdot\AA'$ for
$\AA'\in C(\frac12 \r_0)$.
\*

\item
\0(b) $F_1(\V A')$ =  second order remainder of the $\V
A'$-expansion of $\lis f_0(\V A'+\aa)$:
\be \|F_1\|_{\wt\r,\k_0}\le \wt\g_1 \e_0 \frac{\wt\r^2}{\r_0}
\qquad \V A'\in\CC(\wt\r,\k_0),\quad \wt\r\le\frac12\r_0\Eq{e2.14}
\ee
with $\wt\g_1$ dimensionless. Hence $\|\dpr_{\AA'}
F_1\|_{\wt\r}\le \g_1 \e_0 \frac{\wt\r}{\r_0}$,
$\wt \r<\frac14\r_0,\g_1=4\wt\g_1$.
\*

\item
\0(c) For 
$\wt\r\le\frac14\r_0$ by
Eq.\,\ref{e2.5}\,\ref{e2.12}\,
it is $|F_2|_{\frac12\r_0,\k_0}< \e_04\r_0\g_2\frac{\e_0}C \k_0^{-c_2}$ in
$\CC(\frac12\r_0,\k_0)$. Hence, for $\g_3,c_3$ dimensionless:
\be \eqalign{
&|\dpr_{\AA'}F_2|_{\wt\r,\k_0}
+\frac1{\wt\r}|\dpr_\Ba F_2|_{\wt\r,\k_0}\le \g_3\e_0\frac{\r_0}{\wt \r}
\frac{\e_0}C\k_0^{-c_3}\cr}
\Eq{e2.15}\ee 
on the two contributions to $F_2$ (remark that
$\r_0-\frac14\r_0>\frac14\r_0$ and that only the first is responsible
of the factor $\frac{\r_0}{\wt\r}$).\,\footnote{\small Remark: the domain in
$|\AA'|$ has been reduced to $<\frac14\r_0$ to permit the
dimensional estimate of the derivative of the first of the two
terms in $F_2$, as is done below also to bound $F_3$.}
\*

\item
\0(d) Finally $F_3$, with $\F_0$ defined in Eq.\equ{e2.8}, does not
vanish, but {\it an occurring algebraic cancellation} 
simplifies it, under the conditions in Eq.\ref{e2.10} and, using Eq.\,\ref{e2.9}\,, as:
\be\kern-3mm\eqalign{
   &F_3(\V A',\V z)\defi(\Bo_0\kern-1mm+\kern-1mm J_0 (\V
  A'\kern-1mm+\kern-1mm\aa)\cdot\BDpr_\Ba\F_0\kern-1mm+\kern-1mm
  f_0(\V A'\kern-1mm+\kern-1mm\aa,\Ba)\kern-1mm-\kern-1mm\lis f_0(\V A'+\aa))\cr
  &=-\sum_{\V0\ne \Bn} f_{0,\Bn}(\V A'+\aa)
  \frac{(J_0(\V A'+\aa)\cdot\Bn)^3}{(\Bo_0
  \cdot\Bn)^3}e^{i\Ba\cdot\Bn}\le \wt\g\frac{\e_0}{\r_0}\wt\r^3\k_0^{-c}
\cr} \Eq{e2.16}\ee
and can be bounded in $\CC_{\wt\r,\k_0}$ by
$\wt\g\frac{\e_0}{\r_0}\wt\r^3\k_0^{-c}$ for suitable
dimendionless $\wt\g,c$, and in the variables
$(\AA',\Ba)\in\CC_{\wt\r,\k_0}$ as in c) above:
\be \eqalign{
&|\dpr_{\AA'}F_3|_{\wt\r,\k_0}+\frac1{\wt\r}|\dpr_\Ba
F_3|_{\wt\r,\k_0}
\le \g_4 \e_0
\Big(\frac{\wt\r}{\r_0}\Big)^2\k_0^{-c_4}\cr}\Eq{e2.17}\ee
where $\g_4,c_4$ are dimensionless constants and
$\wt\r\le\frac14\r_0$.
\end{itemize}
\*

Therefore if $\wt \r\le\frac14\r_0,\wt k\le \k_0$ and if
Eq.\ref{e2.10} holds:
\be\eqalign{
&J_1=J_0,
\qquad|\BDpr_{\V A'}F_1(\V A')|\le \lis\g \e_0\Big(\frac{\wt\r}{\r_0}\Big),\cr
&|\BDpr_{\V A'}F_2(\V A',\Ba)|+\frac1{\wt\r}|\BDpr_\Ba F_2(\V
A',\Ba)|\le \lis
\g \e_0\frac{\e_0}C \frac{\r_0}{\wt\r}\k_0^{-\lis c}, \cr
&|\BDpr_{\V A'}F_3(\V A',\Ba)|+\frac1{\wt\r}|\BDpr_\Ba F_3(\V
A',\Ba)|
\le \lis\g  \e_0 \Big(\frac{\wt\r}{\r_0}\Big)^2 \k_0^{-\lis c}\cr
}\Eq{e2.18}\ee
for $\lis\g,\lis c$ constants (dimensionless), $(\V
A',\Ba)\in\CC(\wt\r,\k_0)$. 
\*

Set $F'_j(\AA',\Ba')=F_j(\AA',\Ba'+\D(\AA',\Ba'))$; the size
$\e_1$ of $f_1(\AA',\Ba')$, can be easibly bounded, in
$\CC(\frac14\r_0,\k_0-2\d_0)$ with $\d_0\le \lis\g
(\frac{\e_0}C)(\k_0)^{-\lis c}<\frac12\k_0$, by composing the
$\Ba$-derivatives of the $F_j(\AA',\Ba)$ with the $\AA'$ and
$\Ba'$-derivatives of $\D_j(\AA',\Ba')$. Via Eq.\equ{e2.13} it is
seen that the bounds on $\dpr_{\AA'} F'_j(\AA',\Ba')$ and
$\dpr_{\Ba'}F'_j(\AA',\Ba')$ {\it remain the same} as those on
$\dpr_{\AA'} F_j(\AA',\Ba)$ and $\dpr_{\Ba} F_j(\AA',\Ba)$
provided the constants $\g_j,c_j$, introduced above, are
increased to a suitably larger pair $\lis\g,\lis c$.  See
Appendix\,\ref{B}\,.\,\footnote{\small Remark that $J_{0,\pm},C$
are given and, without loss, $\r_0,\k_0$ can be reduced at will,
so that Eq.\equ{e2.10} is really a condition on $\e_0$.}

Thus the corresponding $F'_j(\AA',\Ba')\defi F_j(\AA',\Ba)$ admit
the same bounds as Eq.\ref{e2.18}, but in the smaller domain
$\CC(\frac12\r_0,\k_0-2\d_0)$ with
$\d_0\le \lis\g(\frac{\e_0}C)\k_0^{-\lis c}<\frac12\k_0$, by
simply suitably increasing the constants $\lis \g,\lis c$, as
discussed in Appendix\ref{B}\,.

Hence the  Eq.\ref{e2.18} considered as estimates of the
derivatives of $F'_j(\AA',\Ba')$ (with the larger
$\lis\g,\lis c$ and $\d_0=\lis\g(\frac{\e_0}C)\k_0^{-\lis c}<\frac12\k_0$)
lead immediately to an estimate of the size
$\e_1=|\dpr_{\AA'} f_1|_{\wt\r,\wt\k}+\frac1{\wt\r}|\dpr_{\Ba'}
f_1|_{\wt\r,\wt\k}$ of the perturbation
$f_1(\AA',\Ba')=\sum_{j=1}^3 F_j(\AA',\Ba'+\D(\AA',\Ba'))$:
namely $\e_1\le C \g ({\frac{\e_0}C})^{\frac32}
\k_0^{-c}$.\,\footnote{\small Remark that here it is necessary
to regard the $F'_j$ as defined in $\CC(\wt\r,\wt\k)$ with
$\wt\k=\k_0-2\d_0$ to make use of dimensional bounds on
$\D(\AA',\Ba')$ implied by Eq.\ref{e2.13}.}\,See
Appendix\,\ref{B} for details.
\*

Given $\r_0,\k_0,n\ge1,\l=\frac12$ define, for
$\lis \g,\lis c$, as in Eq.\ref{e2.18}:
\be \r_n=\Big(\frac{\e_{n-1}}{C}\Big)^{\l}\r_{n-1},\quad \k_n=\k_{n-1}-2\d_{n-1},
\quad \d_n=\lis \g \frac{\e_n}C \k_n^{-\lis c}
\Eq{e2.19}\ee
Iterate the renormalization transformation
$(\AA_{n-1},\Ba_{n-1})\to (\AA'=\AA_n,\Ba'=\Ba_n)$ transforming
the Hamiltonian $H_{n-1}(\AA_{n-1},\Ba_{n-1})$, on
$\CC(\r_{n-1},\k_{n-1})$ with pertubation $f_{n-1}$, into
$H_{n}(\AA_{n},\Ba_{n})$, on $\CC(\r_{n},\k_{n})$ with pertubation
$f_{n}$. Suppose inductively, for all $n\ge0$:
\be \g_0\e_n J_{0,-}^{-1}<\r_n,\qquad C^{-1}\r_n\ J_{0,+}<1,\qquad
\k_n>\frac12\k_0 \Eq{e2.20}\ee
Then, for $(\V A',\Ba')\in \CC(\r_n,\k_n)$, 
the $\e_n=\|\BDpr_{\V A'}
f_n\|_{\r_n,\k_n}+\frac1{\r_n}\|\BDpr_{\Ba} f_n\|_{\r_n,\k_n}$ are
bounded, {\it if}  $\k_n>\frac12\k_0$, by:
\be
\eqalign{
&\frac{\e_n}C=\g \Big(\frac{\e_{n-1}}C\Big)^{\x}
{(\frac{\k_0}2)^{-c}}, \qquad \k_n=\k_{n-1}-\d_{n-1}\cr}\Eq{e2.21}\ee
with $\g,c$  dimensionless, $\x=\frac32$.

The recursion, Eq.\,\ref{e2.19}\,, for
$\frac{\e_n}C$ can be
iterated for $n\ge1$, {\it as long as the conditions Eq.\ref{e2.10}
remain valid}, finding
$\frac{\e_n}{C}\le \L (\frac{\e_{n-1}}C)^{\x}$ for
$\L=\lis\g (\frac{\k_0}2)^{-\lis c}$ and $\g,c,\x=\frac32$
dimensionless constants. If $\h_n\defi \s\frac{\e_n}C$
with $\s=\L^{2}$ this becomes
simply $\h_n\le \h_0^{\x^{n-1}}$, and consequently
$\r_n\le \L^{-\frac{n(n+1)}2} \h_0^{\x^n-1}\r_0=O( 
\h_0^{\x^n})\r_0=O((\L^2\frac{\e_0}C)^{\x^n})\r_0$.

Then Eq.\ref{e2.19} imply that
$\k_n=\k_0-2\sum_{k=1}^n\lis \g \frac{\e_{n-k}}C
(\frac12\k_0)^{-\lis c}$ is bounded by
$\k_n=\k_0-2\sum_{k=0}^{n-1}\L^{-1} \h_{k}$: therefore
the conditions in Eq.\ref{e2.20} can be reproduced if
$\h_0=\s\frac{\e_0}C=\L^2\frac{\e_0}C$ is small enough.

The result is that in the coordinates $(\V A',\Ba')\defi(\V
A_n,\Ba_n)$, up to a time scale $\sim
\Bo_0^{-1}O(\L^{2} \Big(\frac{\e_0}{C}\Big)^{-\x^n})$ and in
region of phase space of order
$\r_n\sim \r_0\Big(\L^2\frac{\e_0}{C}\Big)^{\x^n}$, the motion is
quasi periodic motion with spectrum $\Bo_0$ and Hamiltonian $H_n$
with perturbation size $\sim\e_n\sim(\L^2\frac{\e_0}C)^{\x^{n}}$.

The conclusion is that analyticity in the $\AA$ variables is
lost: in the limit $n\to\infty$ remains a quasi periodic motion
confined for time $\in (0,+\infty)$ on the real (and analytic)
torus to which the approximations $\AA'_n,\Ba'_n$ lead.

\begin{itemize}
\item
It is also possible to define a sequence of maps $\wt \KK_n$
defined in the {\it fixed domain} $\CC_{\frac14\r_0,\frac14\k_0}$
by rescaling the action coordinates of the polydisks by a factor
$(\frac{\e_{n-1}}C)^{-\frac12}=\r_{n-1}/\r_n$, $n\ge1$ so that they are all
turned into $\CC_{\frac14\r_0,\frac14\k_0}$: calling
$\wt\h_n\defi\frac{\e_{n-1}}C$ the rescaling
transformation will rescale time by $\wt \h_n^{-\frac12}$ and will change $\V A_n$ into $\V
A'_n=\wt\h_n^{-\frac12}\V A_n$ and change the Hamiltonian into:
\be\wt H_n=\Bo_0\cdot\V A'_n+ \wt\h_n^{\frac12} \frac12(\V A'_n\cdot
J_0\V A'_n) +\wt\h_n^{-\frac12} f_n(\wt\h_n^{\frac12}\V A'_n,\Ba'_n)
\tende{n\to\infty} \Bo_0\cdot\V A'_{\infty}\Eq{e2.22}\ee 
and in the rescaled variables the sizes of the anharmonic terms
tend to $0$ superexponentially, taking into account the recursion
defined in Eq.\equ{e2.20} which implies that the size of $f_n$ is
of order $\frac{\e_n}C=(\wt\h_n)$.

\item
Hence the perturbation $f_0$ and the twist $J_0$ are,
after renormalization, ``irrelevant'' operators (in
Eq.\equ{e2.17} both tend to $0$ as $n\to\infty$), while the
harmonic oscillator $\Bo_0\cdot\AA$ is a ``fixed point'': in some
sense the transformation has the harmonic oscillator as an {\it
attractive fixed point}. This completes a proof of the KAM
theorem, {\it interpreted in the Renormalization Group frame}
\cite{Ko954,Ar963b,Mo962,BGGS984}: it can be classified as a
``{\it super-renormalizable}'' problem, as it {\it requires only
a second order perturbation analysis around the trivial fixed
point}, Eq.\equ{e2.7}.
\end{itemize}
\*
\0{\it Remarks:}

\0(1) The analysis of this section follows
closely \cite{Ga019b} correcting a few dimensional typos.

It is possible to redefine $J_1\equiv J_0+\frac12\dpr^2_\aa \lis
f(\aa)$ and correspondingly add to $F_1$, in Eq.\equ{e2.5}, the
second order term in $\AA'$ namely
$-\AA'\cdot\frac12\dpr^2_\aa \lis f(\aa)\AA'$. This modifies
$F_0$ in the first line of Eq.\,\ref{e2.5} replacing $J_0$ with
$J_1$ and the improves the bound of $F'_1$ to:
$|F_1'(\AA')|<\g_1\e_0 (\frac{\wt\r}{\r_0})^2$.  Then defining
$\r_n=(\frac{\e_0}{C})^{\l}\r_{n-1}$ with $\l=\frac13$ it
follows, as above, $\frac{\e_n}C< \L(\frac{\e_0}C)^{\frac43}$
(as, in this case, the best choice of $\l$ is $\l=\frac13$ so
that $\x=\frac43>\frac32$).

In the $\AA_n,\Ba_n$ the motion is controlled by a quadratic
Hamiltonian $\Bo_0\AA_n+\frac12 \AA_n J_n\AA_n$ in a portion of
action space $O((\frac{\e_n}C)^{\frac13} \r_0)$ ''much larger''
compared to the previous case, where the size was
$O(\frac{\e_0}C)^{\frac12} \r_0$, but $J_n=J_0$ was simpler.

The implication is that up a time scale of order
$\o_0^{-1} (\frac{\e_0}C)^{-\frac43 n}$ the  motion is quasi
periodic in a domain of size $\r_0 (\frac{\e_0}C)^{(\frac13)^n}$
controlled by Hamiltonian with quadratic part $J_n\ne J_0$.
The previous choice gave control up to a time shorter 
$\o_0^{-1} (\frac{\e_0}C)^{-(\frac32)^n}$ controlled
by the simpler Hamiltonian with quadratic part $J_n=J_0$ but
in a domain of much smaller size $\r_0 (\frac{\e_0}C)^{(\frac12)^n}$.

\0(2) The role of the constant $C$ is essential, of course,
but it should be remarked that it appears in the smallness
condition of $\frac{\e_0}C$ and in $J_{0,+}\r_0 C^{-1}<1$: both
conditions can be fulfilled if $C=O(\e_0^\x)$ with $\x<1$.
Question: could this be of interest (aside the simple application
in the remark in the following Sectiom, see \,\ref{Ceps}.
Sec.\ref{sec3}).

\0(3) The analysis of the first iteration is extremely simplified
if $f_0$ depends only on $\Ba$: this is a {\it Thirring model}
and in Eq.\ref{e2.5} it is $\aa=0$, $F_1=0$, $F_2$ contains only
the quadratic term but $F_3$ still has to be defined via
Eq.\ref{e2.8} so that at the next iteration dependence on $\AA'$
is introduced in $f_1$. Nevertheless the above proof remains
valid and yields existence of the invariant torus with spectrum
$\Bo_0$ {\it without} any condition on the lower bound $J_{0,-}$
as long as it is $>0$. The invariant tori in Thirring models are
called ``twistless'' as no upper bound on $J_{0,-}$ is necessary:
for such models a direct proof based on the explicit computation
of a series, ``Birkhoff series'', converging to the parametric
equations of the invariant torus, can be derived, \cite{Ga994b}\,,
and covers also some cases in which
$J_{0,-}=0$.\,\footnote{\small Which do not fit immediately in
the above proof: the latter can be adapted to the case
$H=\Bo_0\cdot\AA+\frac12 \AA^\perp\cdot
J_0\AA^\perp+f_0(\AA^\perp,\Ba)$ where
$\AA^\perp=(A_1,A_2,\ldots,A_p,0,\ldots0)$ if $J_{0,j}>0$ for
$j\le p$ and $(J_0)_j=0$ for $j>p$, which generalizes
the Thirring models and is even an easy extension of the twistless
Hamiltonians theory (as the perturbation may now depend also on
$\AA^\perp$): the extension is also implied by the analysis
in \cite{Ga994b}\,.}

\def\SEC{FPUT}                     
\section{\SEC}                             
\label{sec3}\iniz                          

The FPUT digital work (reprinted in \cite{Ga008e}), as soon as
the contemporary KAM theorem became widely known, was
qualitatively ``explained'', at least by several
physicists, \cite{Fo992}, by the remark that the observed motions
were generated by a ``small'' perturbation of the simple harmonic
chain and could {\it probably} be treated via the results of the
KAM theory: which, at the same time, was leading to a deeper
understanding of the three body problem. The latter problem, as
well as the FPUT chain, were in fact both perturbations of
integrable systems, although affected by ``resonances''.

However it took a long time to develop a complete proof that low
energy motions observed in some FPUT experiments should mostly be quasi
periodic, \cite{Ri001}.

The models originally chosen in FPUT had Hamiltonian:
\be H=\sum_{j=1}^{N-1}\frac12p_j^2+W_{\a,\b}(q_j-q_{j+1}),\qquad
 W_{\a,\b}(x) = \frac12 x^2 + \a x^3 + \b x^4
\Eq{e3.1}\ee
with {\it fixed endpoints} $q_0=0,q_N=0$, \cite{Fo992}. If $\b=0$
or $\a=0$ the models are respectively called fput-$\a$-chain or
fput-$\b$-chain.

The $\a$-chain, as a dynamical system, presents the difficulty
that $W_{\a,0}(x)$, is unbounded below, if $\a>0$, or above if
$\a<0$: hence it is not even clear whether all or, at least, some
of the motions remain in a bounded phase space region (in a
finite or infinite time). The problem does not arise in the
$\b$-chains if $\e>0$ as it is always supposed here.

In the original FPUT work nine choices of $\a$ and $\b$ are considered
and the cases numbered $4,5,6,7,9$ are apparently $\b$ models
(\ie $\a=0,\b>0$) while those numbered $1,2,3,8$ are $\a$ models
($\a>0,\b=0$). In all simulations the motions kept a total energy
constant, ``within $1\%$ or so''.


The moving particles coordinates, ${N}-1$ in number, can be
expressed as $q_k=\frac{\sqrt2}{\sqrt N}\sum_{h=1}^{N-1}
\sin(\frac{\p h k}{N}) Q_h$
and $p_k=\frac{\sqrt2}{\sqrt N}\sum_{h=1}^{N-1}
\sin(\frac{\p h k}{N}) P_h$ for $k=0,\ldots,N$.
\*

The well known theory of the vibrating string with elastic
potential energy $\sum_{k=0}^{N-1} \frac12
(q_k-q_{k+1})^2=\frac12\sum_{h=1}^{N-1}\o_h^2 Q_h^2$ induces to
study its oscillations via Forurier's transforms
coordinates $(\bf P,\bf Q)$  expressed as functions of
$(\pp,\qq)$ to which  are
canonically conjugated.  If $\o_h=2\sin\frac{\p
h}N$, therefore:
\be H_0(\pp,\qq)= H_0({\bf P,\bf Q})=
\sum_{k=1}^{N-1}\frac12 (P_k^2+\o_k^2
Q_k)^2\,\defi\,\sum_{k=1}^{N-1}\o_k a_k
\Eq{e3.2}\ee
where $a_k=\frac{(P_k^2+\o_k^2 Q_k^2)}{2\o_k}$. The latter is a
natural coordinate as the pairs
$(a_k,\f_k)= (a_k,\arcsin(\frac{\o_k Q_k}{\sqrt{2\o_k a_k}}))$
are {\it action-angle} variables and
evolve, in the elastic chain, as
$(a_k,\f_k)\to (a_k,\f_k+\o_kt)$; and 
$(\pp,\qq)\otto ({\bf P,\bf Q})\otto (\aa,\Bff)$ are symplectic
maps.

Therefore $Q_k=(\frac{2a_k}{\o_k})^{\frac12} \sin\f_k$ and, as an example,
the fput-$\b$-chain Hamiltonian can be written,
if $s_h=\sin \frac{\p h}N,c_h=\cos\frac{\p h}N$:
\be \eqalign{
&H_\e(\pp,\qq)= \sum_{k=1}^{N-1} \o_k \a_k+
\e\sum_{\hh} T_\hh \prod_{i=1}^4
Q_{h_i}=\Bo_k\cdot\aa+\e f_0(\aa,\Bff)\cr
&T_\hh=(\frac2{N})^{2}\sum_{k=1}^{N-1} \prod_{j=1}^4
(s_{h_jk}(1-c_{h_j})-c_{h_jk}s_{h_j}),
\cr
&f_0(\aa,\Bff)=\frac1{2^4}\sum_{\hh\in Z^4,\Bs}
T_\hh \prod_{j=1}^4 \s_j e^{i\s_j\f_{h_j}}
(\frac{2 a_{h_j}}{\o_j})^{\frac12},
\cr}\Eq{e3.3}\ee
where $\Bs=(\s_j)_{j=1}^4, \s_j=\pm1$ and
the variables $Q_k$ have been expressed in terms of the
$\aa,\Bff$.
\*

The first mathematical results, \cite{Ni971}, dealt with the
$\b$-chain with fixed extremes $q_0=q_N=0$, considered as a
perturbation of order $\e$ of the elastic chain with $N-1$
particles (\ie of the chain with $\a=0,\b=0$).
\*

Perturbation theory can be applied to $H_\e$: just look for a
canonical transformation generated by a function $\e \F(\aa',\Bff)$
as $\aa=\aa'+\e\BDpr_{\Bff}
\F(\aa',\Bff),\Bff'=\Bff+\e\BDpr_{\aa'}\F(\aa',\Bff)$ which, if
possible, transforms $H_\e(\AA,\Bff)$ into $H_\e(\aa',\Bff)$ (where
$\Bff$ is imagimed expressed as function of $(\aa',\Bff')$).
\be H_\e(\aa',\Bff)=\Bo\cdot\aa'+\e\Bo\cdot\BDpr_\Bff \F(\aa',\Bff),\Bff)
+\e f_0(\aa'+\e\BDpr_{\Bff}\F,\Bff)\Eq{e3.4}\ee
Naturally $\F$ would be chosen so that $H_\e(\aa',\Bff)$ is
$\Bo\cdot\aa'$. Calculating up to $O(\e^2)$ this means that $\F$
should be such that $\e\Bo\cdot\BDpr_\Bff \F(\aa',\Bff) +\e
f_0(\aa',\Bff)$ vanishes.
However even this might be impossible if there are $\Bn\in Z^{N-1}$ such
that $\Bo\cdot\Bn=0$.

Therefore the best that can be done is to subtract from the above
$f_0$ the terms which, expressed in the Fourier's series, have
``resonant'' harmonics (orthogonal to $\Bo$ but $\ne 0$) and
define $\F$ to cancel, at lowest order, the part of $f_0$
left. So $\F$ can be chosen so that the new Hamiltonian (once
laboriously expressed in the variables $(\aa',\Bff')$) will
contain, up to order $\e^2$, only 'resonant harmonics'
(including $\Bn=0$).

The new Hamiltonian will have the form
$H_\e(\aa',\Bff')=\Bo\cdot\aa'+\e \lis H_1(\aa',\Bff') +\e^2
H_2(\aa',\Bff')$ where $\lis H_1$ is entirely resonant.

Repeating the argument $s$-times the Hamiltonian is represented
by $H_\e=H_2+\sum_{k=1}^s\e^s\lis H_{s} + \e^{s+1}H_{\e,s+1}$
where the $\lis H_s$ contain only resonant terms, which remain
the same if $s$ is increased. The series obtained in this way is
the ``Birkhoff series'', which however is generically divergent as
shown by Poincar\'e, \cite{Ga993}\,.

However truncating the sum to an order $s\ge1$ and neglecting the
$\e^{s+1}H_{\e,s+1}$ might turn out to be a simpler system. This
happens in the present case if $N$ is prime or is not a power of
$2$, as pointed out in \cite{Ni971}, simply because there are no
$\Bn\ne0$ with $\Bo_0\cdot\Bn=0$ in the Fourier's expansion of
the potential (\ie no resonances $\Bn$ with $|\Bn|\le4$). The
terms of order $\le \e^2$ in $\lis H_1,\lis H_2$ not only {\it do
not depend} at all on the angles $\Bff'$, but $\lis H_1=0$ and
the terms in $\lis H_2$ can be collected in
the form $\e^2 \aa'\cdot J \aa'$, where $J$ is explicitly computed
in terms of the spectrum $\Bo_0$ of the elastic chain and
furthermore it is $\det J\ne0$.

Hence the Hamiltonian $\lis H_\e=\Bo\cdot\aa'+\e^2 \lis H_2(\aa')$
is integrable and, as computed in \cite{Ni971}, is found to be:
\be \lis H_\e(\aa')=\sum_{k=1}^{N-1}\o_k a'_k+
\frac{\e^2}{2N}\Big( \sum_{h,k}^{1,N-1} \frac{\o_h\o_k}8\ a'_h
 a'_k-\sum_k\frac{\o_k^2}{32} {a'_k}^2\Big)
\Eq{e3.5}\ee
The surprising simplicity of this normal form, and the $\o_h\ne0,
h=1,\ldots,N-1$, immediately allows to check that
$\Bo\cdot\aa+ \frac\e2 \aa\cdot J\aa$ has a Jacobian matrix $J$
constant and $\det J\ne0$.

The {\it few} values of $N$, \ie other than a prime or a power of
$2$, which were still missing has been solved in \cite{Ri001}
where a careful analysis shewed that no restriction should be set
on the $N$ of fput-$\b$-chains with fixed extremes, and
identified the basic reason for the existence of an integrable
Birkhoff transform permitting to avoid the issue of resonances
affecting non trivially the Birkhoff transform, see
also \cite{Ri008}\,.

The difficulty remaing after \cite{Ni971} was that, given $N$, it
was not excluded that resonant harmonics were absent in the
Fourier expansion of $H_2$: the only way out seemed to compute
the corresponding Fourier's coefficients. A task that could be
performed (as it was done as time elapsed) also for single
choices of $N$, but not covering all values of $N$.

The breakthrough idea, \cite{Ri006}, is to {\it forget}
(temporarily) the computation of the normal form and to check
that the resonant harmonics, that could contribute resonances to
the normal form, could not be present because they break
symmetries of the Hamiltonian (which are preserved in the normal
form construction). Thus the normal form can be constructed
correctly as if the assumption in \cite{Ni971} could be valid
(although it was known to be false in few interesting cases: for
examples (with $N=15,21$) see \cite[Sec.8.5]{Ri008}\,. The
symmetries are briefly presented in Appendix\ref{C}.

The spectra of the motions of $\lis H_\e(\aa')$ will have the
form $\Bo=\Bo_0+\e J\aa'$\label{Ja} and $KAM$ can be applied to
perturbations of the considered integrable model. Let $\O$ be an
open region in phase space, in which the $\lis H_\e(\aa')$ is
defined: then $\O$ is covered by the tori corresponding to the
actions $\aa$: \eg $0<H_\e(\aa')<E$.

The $\det J\ne0$ implies that the set in $\O$ of invariant tori
 for whose spectra verify the inequality $|\Bo\cdot\Bn|> C |\Bn|$
 fails, for some $0\ne\Bn\in Z^N$, has volume $<C\e^{-2} \k
 |\O|$, for a constant $\k>0$: hence the for almost points in
 $\O$ the spectrum $\Bo$ verifies a Diophantine inequality.

Hence under a perturbation of $\lis H_\e$ of size $\h$ a large
part of phase space will remain covered by invariant tori quasi
periodically run with $C_\e=C\e^2$-Diophantine frequencies if
$C_\e$ is large enough compared to $\h$. The $\b$-chain being
approximated to order $\h=\e^2$ by the above integrable system
$\lis H_\e(\aa')$ will therefore have a large part of a set $Q$
in phase space covered by $C_\e=C\e^2$-Diophantine tori. In
particular this applies to the set $Q$ of energy close to the
ground state of the linear chain where $H(\pp,\qq)<E$ (if $E>0$
and $\e$ is small enough), and the fraction of $Q$ covered by
invariant tori will approach $1$ as $\e\to0$ at fixed $E>0$.
\*
\0{\it Remark:} (1) The statement that KAM implies that at small energy
$H_\e(\pp,\qq)<E$ ``most phase space points are on invariant
tori'' requires some care.  Describing the system in the
coordinates $(\aa',\Bff')$ the set $\lis H_\e(\aa')$ depends on
$\e$: but the application of KAM to $\lis H_\e(\aa')$ imposes a
condition on the same $\e$. The remark \label{Ceps} at the end of Sec.2 is
useful here, as it stresses that if the KAM theory is applied to
cases in which the Diophantine constant has the form $C \e^{-1}$
the KAM theorem holds provided the perturbation size $\h\e^{-1}$
with $h$ small enough.
\*

\0(2) About the $\a$-model it is possible that, applying
perturbation theory to the fput-$\a$-chain, an equivalent
Hamiltonian is obtained which is an $O(\e^2)$ perturbation of an
integrable Hamiltonian with $J_\e$ still of $O(\e)$: but this is
not (yet?) proved, \cite{Ri008}. See also the heuristic analysis in
Section\ref{sec4}.
\*

\0(3) The approach to the $\b$-model at fixed extremes can
be applied to consider the Toda lattice chains with periodic or
fixed boundary conditions as approximations to the
fput-$\b$-chains models. The above analysis, \cite{Ri001},
suggests to try to use integrability of the Toda lattice to play
the pertubation theory role in the search of an integrable
Birkhoff approximation.

The suggestion is
to consider the fput-Hamiltonians as perturbations
of the Toda-lattice and consequences of integrability of several
forms of its Hamiltonian:
\be \eqalign{
&H(\pp,\xx)=\sum_{j=1}^{N-1}\frac12p_j^2+
\sum_{j=0}^{N-1}W_T(q_j-q_{j+1})\cr
&W_T(x)=\frac1{\e^2} e^{\e x}=\frac1{\e^2}(1+\e x)+\frac1{2!}
x^2+\frac{\e}{3!}x^3+\frac{\e^2}{4!}x^4+
\frac{\e^3}{5!}x^5+\ldots\cr}
\Eq{e3.6}\ee
integrable for instance with boundary conditions $q_0=q_N$, {\it
periodic} lattice, \cite{Fl974,He974,HK006}, or
$q_0=-\infty,q_N=+\infty$, lattice with {\it endpoints fixed at
$\pm\infty$}, \cite{Mo975b}, or with fixed endpoints $q_0=q_N=0$.

In a sense the idea is that the Toda chains might play the role of
an integrable Birkhoff normal form, of which fput-chains can be
considered a perturbation.  In the next section
the basic
results, \cite{HK006}, on the normal form of the periodic
Toda-chains are summarized and heuristically applied to the
fput-chains.
\*

\def\SEC{Toda and FPUT: heuristics}
\section{\SEC}                             
\label{sec4}\iniz                          

The last suggestion in Sec.\ref{sec3} is to try to take advantage of the
very detailed studies on the Toda chain described below.
For this purpose a periodic Toda Hamiltonian will be
written, \cite{HK006}:
\be H(\pp,\qq)=\sum_{k=1}^N\Big(
\frac12 p_k^2+\a^2e^{q_k-q_{k+1}}\Big)=\sum_{k=1}^N
(\frac12 p_k^2+\a^2V_T(q_k-q_{k+1}))
\Eq{e4.1}\ee
with $q_1=q_{N+1}$. This is obtained by a canonical
transformation on Eq.\ref{e3.2}: $p=\e p'$, $q=\frac1\e q'$
followed by a rescaling of time $t'=\e^2 t$ leading to
Eq.\ref{e3.4}, where $\a$ is a new coupling equal to $\e^{-2}$.

The {\it periodic} Toda lattice model is studied
 in \cite{HK006} where $2N$ global canonical action-angle
 variables $(\aa,\Bff)\in (R_+^N,[0,2\p]^N)$ are constructed and
 evolve at time $t$ into $(\aa,\Bff+\Bo(\aa)t)$, following a
 Hamiltonian $H_\a(\aa)$. Near $\aa=0$ it is
 $\o_i(\aa)=2\a \o_{0,i}+ \frac1{4N}a_i +O(|\aa|^2)$, if
 $\Bo_0$ is the spectrum of the linear oscillations
 $\o_{0,i}=2\sin \frac{\p i}N$.

The $H_\a(\aa)$, \cite[th.1.4]{HK006}, is {\it globally defined}
 for $\aa\in R^N_+$ and the Jacobian
 $J_\a(\aa)=\det\frac{\dpr \Bo(\aa)}{\dpr \aa}$, \ref{Ja}, is analytic in
 $\a>0$; for $\a>0$ it is strictly convex in $\aa\in R_+^N$ and
 $J_\a(\aa)_{ij}=\frac1{4N}\d_{ij}+O(|\aa|)\ne0$ and the spectrum
 is $\o_i=2\a \sin(\frac{i\p}N)+\frac1{4N}a_i+O(|\aa|^2)$.

The general KAM theory of Sec.\ref{sec2} can be applied, as in
Sec.\ref{sec3}, to perturbations of the Toda-model because
the Jacobian $J_\a$ strict convexity for $\a$ in any interval
bounded away from $0$ implies that the spectrum has the
Dipohantine property: so pertubations of size $\h$ will leave
most of the invariant tori only slightly deformed if $\h$ is
small.

However application to the fput-chains requires more analysis
mainly because the use of the change of variables leads to the
Hamiltonian in Eq.\ref{e4.1} with $\a$ small and compare it with
Eq.\ref{e3.6} $\e$ small: aside from the time rescaling, this
means to compare motions with $\a$ in Eq.\ref{e4.1} to motions
with $\e$ in Eq.\ref{e3.6} setting $\a=\e^{-2}$; so when $\a$ is
very large.

The analysis in \cite{HK006} is global and the coordinates
$(\aa,\Bff)$ exist for all $\a>0$ and transform analytically the
Hamiltonian Eq.\ref{e4.1} into a function $\wt H_\a(\aa)$ with
convex Jacobian $J_{ij}=\frac{\dpr^2_{a_i} H(\aa)}{\dpr^2 a_j}$,
see th.(1.1),th.(1.2) in \cite{HK006}. Hence in the region of
couplings relevant for the fput-chains, \ie $\e$ close to $0$,
$\a=\e^{-2}$ large, the Toda lattice Hamiltonian admits
action-angles coordinates.

Proceeding {\it heuristically}: the Toda lattice can be
considered a perturbation of order $\e^3$ of the
fput-lattice $H_{3,4}$, see Eq.\ref{e3.6}:
\be H_3(\pp,\qq)=\sum_{j=0}^{N-1}
\Big(\frac12 (p_k^2+(q_k-q_{k+1})^2)+\frac\e{3!}
(q_k-q_{k+1})^3+\frac{\e^2}{4!}
(q_k-q_{k+1})^4\Big)\Eq{e4.2}\ee
Hence the periodic elastic chain perturbed by
$h_{3,4}=\frac\e{3!}\sum_{j=0}^{N-1} (q_k-q_{k+1})^3+
\frac{\e^2}{4!}\sum_{j=0}^{N-1} (q_k-q_{k+1})^4$ and the
periodic Toda lattice truncated to order $\e^2$ are the same:
thus their first two orders Birkhoff's normal forms, to remove
the resonant terms present in $H_{3,4}$, coincide up to the next
order above $\e^2$ (the first steps in the construction of
Birkhoff normal form only involves the common part
$h_{3,4}$). The integrability of the Toda lattice should then
imply that the normal form for the above truncation is integrable
implying that the normal form of the above fput-chain is
integrable to order $\e^2$.

Therefore the argument of \cite{Ri001} can be applied
(heuristically here, and avoiding any explict calculation of the
Birkhoff normal form) to reach the same conclusion and the
periodic fput-chain model in Eq.\ref{e4.2} shows quasi periodic
motions at small $\e$, filling
large regions of phase space in any region $\O$ where $e<H<E$, up
to a set of volume approching $0$ as $\e\to0$.

The chain considered in Eq.\ref{e4.2}, $H_{3,4}$, has a
Birkhoff's normal form which to order $\e^2$ is the sum of the
(resonant) contributions to the $\a$-chain (to second order) and
those to the $\b$ chain (to first order): making use of the
arbitrariness of $\e$ it follows that both $\a$ and $\b$ chains
have an integrable normal form, to order $2$ included, and the
KAM theory can be applied to both, if the Jacobian $J$ \ref{Ja} is
non singular (as it is in the Toda lattice and in the
$\a$-model, \cite{HK006,Ri001}).

\centerline{\large\bf {Appendices}}
\APPENDICE{1}
\def\SEC{Shift equation}
\section{\SEC}                             
\label{A}\iniz                          

\0
Existence of the shift $\aa=-J_0^{-1}\dpr_\aa \lis f_0(\aa)$.
\\ 
Remark the bound on the matrix elements
$\s_{ij}=J_0^{-1}\BDpr_{\aa}^2\lis f_0(\aa)$ in $C_{\l\r_0}$ by
$\s=((1-\l)\r_0)^{-1}J_{0,-}^{-1}\e_0 $: hence the map
$S\aa=\aa+J_0^{-1}\BDpr_\aa\lis f(\aa)$ has non singular Jacobian
if $N\s<\frac13$ (for instance)\,
\footnote{\small\label{Jacobian} Let $|v|=\sum_i|v_i|$ for $v\in C^N$. Then the
matrix $\d_{ij}+\s_{ij}$ applied to $v\in C^N$ results
in $w_i=v_i+\sum_j\s_{ij}v_j$: \ie $|w|=\sum_i|w_i|\ge
|v|(1-N\s)>0$ if, say,  $N\s<\frac13$.}\, and the map is locally
invertible; furthermore any pair $\aa',\aa''\in C((1-\l)\r_0)$ is
mapped into a pair $S\aa',S\aa''$ at distance $|S\aa'-S\aa''|>
(1-\s N)|\aa'-\aa''|$, showing invertibility of $S$ in
$C_{(1-\l)\r_0}$ with image in $C((1+N\s)(1-\l)\r_0)\subset
C(\r_0)$ if $\l$ is fixed $\l=\frac23$ and $N\s<\frac13$.

As $\aa$ varies on the boundary of $C_{(1-\l)\r_0}$ the image
$S\aa$ describes a set surrounding the polydisk of radius
$(1-N\s)(1-\l)\r_0=\frac29\r_0$: therefore the images $S\aa$
cover a set containing $0$ and imply existence of $\aa$ with
$S\aa=0$ and therefore, $|\aa|=|J_0^{-1}\dpr_\aa\lis f_0(\aa)|<
J_{0,-}^{-1}\e_0$. In particular if $\l=\frac23,\g_0=3N$:
\be 3N J_{0,-}^{-1}\e_0\r_0^{-1}<1 \tto
|\aa|<J_{0,-}^{-1}\e_0<\frac1{\g_0}\r_0, \qquad \aa\in
C(\frac1{3N}{\r_0}) \Eq{a.1}\ee
in general any other $\l<1$ would be acceptable provided
$\g_0=N(1-\l)^{-1}$; the choices of $\l,\g_0$ are
just convenient for later use and far from optimal.
\*

\APPENDICE{2}
\def\SEC{Renormalized H: dimensional estimates }                     
\section{\SEC}                             
\label{B}\iniz                         
\let\wh=\widehat

Remark that in
$\CC(\frac12\r_0,\k_0)$ the Jacobian of the map
$\Ba'=\Ba+\dpr_{\AA'}\F_0$ is $(I+\BDpr_\Ba\BDpr_{\AA'}\F_0)$
and, from the bound on $\F_0$, Eq.\ref{e2.12}\,, 
$|\BDpr_\Ba\BDpr_{\AA'}\F_0|<6\wh g \frac{\e_0}C \,\k_0^{-\wh c}$.\,
\footnote{\small Recall that $\F_0$ is bounded
in $\CC(\frac23\r_0,\k_0)$ and $|\aa|<\frac16\r_0$; then use
$\frac23-\frac12=\frac16$.} By Eq.\ref{e2.12},
$|\dpr_{\AA'}\F_0(\AA',\Ba)|<\wh \g\frac{\e_0}C \k_0^{-\wh c}$
and the Jacobian determinant will be $\ge
1-\d_0>0$ if $\d_0\le N!6\wh g \frac{\e_0}C \,\k_0^{-\wh c}$ is
smaller than $1$: for definitness suppose
\be \d_0=N!6\wh g \frac{\e_0}C \,\k_0^{-\wh
c}<\frac12\k_0<1.\label{barg}
\Eq{b.1}\ee

Hence, for such $\d_0$, the
$\Ba'=S\Ba=\Ba+\BDpr_{\AA'}\F_0(\AA',\Ba)$, maps (by
Eq.\,\ref{e2.12})\,, $\CC(\frac12\r_0,\k_0-\d_0)$ into a set
$\subset \CC(\frac12\r_0,\k_0)$ {\it covering} the polydisk
$\CC(\frac12\r_0,\k_0-2\d_0)$ and is also locally invertible (as
its Jacobian is nonsingular) and globally one-to-one, hence holomorphic.\,
\footnote{\small\label{canonical} Remark that,
for the purpose of solving the implicit functions problems in
Eq.\ref{e2.4}\,, an argument can be used similar to the one for
the solution of $\aa=-J_0^{-1}\BDpr_\aa\lis f_0(\aa)$ in
Appendix\,\ref{A}\,. Any pair $\Ba',\Ba''\in C(\frac12\k_0)$ is
mapped, if $|v|=\sum_i|v_i|, v \in C^N$, into a pair
$S\Ba',S\Ba''$ at distance $|S\Ba''-S\Ba'|> (1-
N\|\dpr_{\Ba}\dpr_{\AA'}\F_0\|_{\frac12\r_0,\k_0})\,|\Ba'-\Ba''|$;
hence the map $S$ is one-to-one globally and analyticity of
$S^{-1}$ will also follow from the positivity of the
Jacobian. For a more general and systematic study of the
inversion of the maps in Eq.\ref{e2.4} see Appendicx G
in \cite{Ga008}.} $S^{-1}$ will be denoted
$\Ba=\Ba'+\D(\AA',\Ba')$ and $\D$ inherits the bound
Eq.\ref{e2.13}
$\|\D\|_{\frac12\r_0,\k_0-2\d_0}<\lis\g\frac{\e_0}C \,\k_0^{-\lis
c}$ provided the conditions in Eq.\equ{e2.10} hold together with
$\d_0<\frac12\k_0$, and the constants $\lis \g,\lis c$ are chosen
large enough.\,: its inverse in $\CC(\frac12\r_0,\k_0-2\d_0)$
will be written $\Ba=\Ba'+\D(\AA',\Ba')$ and fullfills the
bound \Eq.\,\ref{e2.13}\, for suitable $\lis\g,\lis c,\d_0$.
\*

For the purpose of dimensional bounds it is convenient to
consider the second map $S$ in Eq.\equ{e2.4} in the domain
$\CC(\frac12\r_0,\k_0-2\d_0)$ with 
$\d_0$ in Eq.\ref{b.1} and values in
$\CC(\frac12\r_0,\k_0-\d_0)$.

The conditions for the bounds are summarized as, Eq.\,\ref{e2.10}\,\ref{b.1}:
\be \g_0\e_0J_{0,-}^{-1}<\r_0,\qquad C^{-1}\r_0\ J_{0,+}<1,\qquad
\d_0=\wt\g \frac{\e_0}C \k_0^{-\wt c}<\frac12\k_0\Eq{b.2}\ee
for $\g_0,\wt\g,\wt c$ suitable dimensionless
constants.\,\footnote{\small\label{likewise}Likewise the
$\AA=\AA'+\aa+\dpr_\Ba\F_0(\AA',\Ba)$ would lead to
$\AA=\AA'-\aa+\X(\AA',\Ba')$ with
$\X(\AA',\Ba')=-\dpr_\Ba\F_0(\AA',\Ba'+\D(\AA',\Ba'))$ under
similar conditions; but $\X$ is not needed here.}

Given a generic function $F(\AA',\Ba)$ define $F'(\AA',\Ba')\defi
F(\AA',\Ba'+\D(\AA',\Ba'))$.  Then, by Eq.\ref{e2.20}, to estimate the
derivatives $\dpr_{\AA'}$ and $\dpr_{\Ba'}$ of $F_j$ consider
first the $\dpr_{\AA'}$ and $\dpr_{\Ba}$. Then make use of, see
also Eq.\,\ref{e2.10}\,\ref{e2.13}\,:

\be\eqalign{
\dpr_{\AA'} F'(\AA',\Ba')=& \dpr_{\AA'} F(\AA',\Ba)+\dpr_\Ba
F(\AA',\Ba)\cdot\dpr_{\AA'}\D(\AA',\Ba')\cr
\dpr_{\Ba'}
F'(\AA',\Ba')=&\dpr_{\Ba}F(\AA',\Ba)\cdot(I+\dpr_{\Ba'}\D(\AA',\Ba'))\cr
\|\D\|_{\frac12\r_0,\k_0-2\d_0}<& \lis\g\frac{\e_0}{C}\k_0^{-\lis
c}\cr
|\dpr_{\AA'}\D|_{\frac14\r_0,\k_0-2\d_0}&+
\frac{4}{\r_0}|\dpr_{\Ba'}\D|_{\frac14\r_0,\k_0-2\d_0}<\lis\g\frac{\e_0}{C}\k_0^{-\lis c}
\cr}\Eq{b.3}\ee

Hence $f_1(\AA',\Ba')=f'_1(\AA',\Ba)\equiv\sum_{j=1}^3
F'_j(\AA',\Ba)$ and Eq.\ref{e2.19}:
\be
\eqalign{&|\dpr_{\AA'} f_1(\AA',\Ba')|=|\dpr_{\AA'}f'_0(\AA',\Ba)
+\dpr_{\Ba}f'_0(\AA',\Ba)\dpr_{\AA'}\D(\AA',\Ba')|\cr
&\le\lis\g \e_0(\frac{\e_0}C)^{\frac12}\k_0^{-\lis c}+
\wt\g\r_0 \e_0(\frac{\e_0}C)^{\frac12}\k_0^{-\lis c}
\frac1{\r_0}\lis\g\frac{\e_0}{C}\k_0^{-\lis
c}\cr
&|\frac1{\wt\r}\dpr_{\Ba'} f'_1(\AA',\Ba')|=
|\frac1{\wt\r}\dpr_{\Ba}f'_0(\AA',\Ba)
(I+\dpr_{\Ba'} \D(\AA',\Ba'))|\cr
&\le |\frac1{\wt\r}\dpr_{\Ba}f'_0(\AA',\Ba)|(1+ N\lis\g\frac{\e_0}{C}\k_0^{-\lis c})
}\Eq{b.3}\ee
Hence the $\e_1$ can be bounded as the $\e'_1$ in Eq.\ref{e2.20}
in the domain $\CC(\frac14\r_0,\k_0-4\d_0)$ (smaller than
$\CC(\frac12\r_0,\k_0-2\d_0)$ to allow using dimensional bounds
in the bounds of the derivatives of $\D$) simply by replacing the
constants $\g_j,c_j$ by a single pair $\lis\g,\lis c$.

\APPENDICE{3}
\def\SEC{FPUT symmetry}                     
\section{\SEC}                             
\label{C}\iniz                          
\def\qq{{\bf q}}

The Hamiltonian
$H_0(\pp,\qq)=\sum_{k=1}^N \frac12(p_k^2+(q_k-q_{k+1})^2)$ for
$N=2(n+1)$ particles can be solved by changing variables
$\qq\otto\wh\qq,\pp\otto\wh\pp$ with:
\be q_h=\frac1{\sqrt N}
\sum_{k=1}^N e^{\frac{2\p i}{N}hk}\wh q_k,\quad p_h=\frac1{\sqrt N}
\sum_{k=1}^N e^{\frac{2\p i}{N}hk}\wh p_k\Eq{c.1}\ee
with $\wh q_h=\lis{{\wh q}}_{N-h}$ and $\wh p_h=\lis{ {\wh
p}}_{N-h}$ and periodic boundary $\wh q_N=\wh q_0, \wh p_0=\wh p_N$.

It becomes
$H_0(\wh\pp,\wh\qq)=\sum_{k=1}^{N} \o_k a_k$
where $\o_k=2\sin\frac{\p k}N$ and
$a_k=\frac12(|\wh p_k|^2+\o_k^2 |\wh q_k^2|)$.
The coordinates $\wh q_N,\wh p_N$ describe the free motion of the
center of mass; discarding $\wh q_N,\wh p_N$ the Hamiltonian is
reduced to $2n=N-2$ degrees of freedom.

It is convenient to separate real and imaginary part of the
$\wh \qq$ and $\wh \pp$ calling $Q_h=-\wh q_{h,im}$ for $h\le\frac{N}2$, and
$Q_{\frac{N}2+h}=\wh q_{h,re}$. Then 
\be q_k=
\frac{\sqrt2}{\sqrt N}\sum_{h=1}^{N/2 -1}
(Q_h \sin\frac{2\p k h}N +Q_{\frac12N+h}\cos\frac{2\p k
h}N)\Eq{c.2}\ee
and define similarly $P_h$ in terms of the $\wh\pp$: in this way
the phase space coordinates for $H_0$ will be
denoted ${\bf Q}=(Q_1,\ldots,Q_{\frac12 N},Q_{\frac12
N+1}\ldots,Q_{N})=({\bf Q_+},{\bf Q_-})$ and, likewise $\bf
P=({\bf P_+},{\bf P_-})$.

Define the canonical maps $S,R$:
\be S\Big(\eqalign{
&{\bf Q_+},{\bf Q_-}\cr
&{\bf P_+},{\bf P_-}\cr}\Big)=
\Big(\eqalign{
&{\bf Q_+},-{\bf Q_-}\cr
&{\bf P_+},-{\bf P_-}\cr}\Big), \qquad
R\Big(\eqalign{
&\qq\cr
&\pp\cr}\Big)=
\Big(\eqalign{&\qq^t\cr
&\pp^t\cr}\Big)
\Eq{c.3}\ee
where $\qq^t=(q_2,\ldots,q_{N-1},q_1)$ if
$\qq=(q_1,\ldots,q_{N-1})$ and likewise is defined $\pp^t$; the
map $S$ is written for the variables $\bf Q$ to make clear that
$S{\bf Q}={\bf Q}$ means, by Eq\equ{c.2}, that the manifold $F$
in phase space of the fixed points of the linear symplectic map
$S$ consists of the ${\bf Q_-}=0,{\bf P_-}=0$ (\ie of the
$\pp,\qq$ with purely imaginary Fourier's tranforms
$\wh\pp,\wh\qq$). And from Eq\ref{c.2} it is seen that on $F$ it is
$q_N=q_0=0$: furthermore the maps $R,S$ commute with $H_\e$
therefore evolution generated by $H_\e$ leaves invariant the
manifold $E$.

Hence the $N$-degrees of freedom periodic b.c. Hamiltonian $H_\e$
describes, if restricted to $F$, a free chain with fixed
endpoints and $n$ degrees of freedom if $N=2(n+1)$. Furthermore
its Hamiltonian can be written:
\be H_0=\sum_{k=1}^n \o_k a_k,\quad a_k=\frac12
\frac{(P_k^2+\o_k^2 Q_k^2)}{\o_k},\quad \sin \f_k=\frac{\o_k Q_k}{\sqrt{2\o_ka_k}}
\Eq{c.4}\ee
with $(a_k,\f_k)_{k=1}^n$ pairs of conjugate symplectic
coordinates $(\aa,\Bff)$.

The relation between the periodic chain, with $N$ particles, and
the fixed extremes chain, with $n=\frac{N}2-1$ particles, leads
to determine classes of integers $N$ for which $N$-particles
chains Birkhoff's series of some low order are integrable at
small $\e$, \cite{Ri001}, including various cases beyond
those with $N$=power of $2$ or  $N$-prime.

Soon, \cite{Ri006}, an original approach was developed to cover
all chains with $N$ arbitrary integer (which by the time had
still remained to relatively few integers): and the result was
achieved without the need to compute the normal form, but using
the symmetries, to infer in general, that fput-$\b$-chains normal
forms could not be affected by possible resonances that, altough
possible, were conjectured not to be involved in the normal form
construction.

It is interesting to remark that the the potential of
fput-$\b$-chain with fixed extremes, setting 
$\Bs=(\s_i)_{i=1}^4, \s_i=\pm1$ and $T_\hh
=(\frac2N)^{2}\sum_{k=1}^{\frac{N}2}
\prod_{j=1}^4((1-c_{h_i})s_{h_ik}-s_{h_i}c_{h_ik})$ for
$\hh=(h_i)_{i=1}^4$, where $c_{h}=\cos\f_{h},s_{h}=\sin\f_{h}$. can be written:
\be V_4({\bf Q})=\frac14\sum_{\hh,\Bs} T_\hh \prod_{j=1}^4
\s_j e^{i \s_j \f_{h_j}}
(\frac{{a_{h_j}}}{{\o_{h_j}}})^{\frac12}\Eq{c.5}\ee
and easily identifies the terms which need to be absent to obtain
a normal form which is integrable: if
$n_{\hh,\Bs}(k)=\sum_{h_j=k}\s_j$ and
$\nn=(n_{\hh,\Bs}(k))_{k=1}^{N=1}$ the uneliminable resonances
$\Bn$ are the ones for which $\x=\nn\cdot\Bo=0$. Hence to prove
the result it is necessary to show that the sum of the terms
labeled by the above $\nn$ for which $\x=0$ either do not arise
(but they do in interesting cases) or that they add up to $0$ as
a consequence of the symmetry or because $\nn=0$, as found
in \cite{Ri006,Ri008}.
\*
STATEMENT: This paper has no supplementary data
\*



\bibliographystyle{unsrt}
\bibliography{0Bib}

\begin{thebibliography}{10}

\bibitem{Bo868-a}
L.~Boltzmann.
\newblock Studien {\"u}ber das gleichgewicht der le\-bendigen kraft zwischen
  bewegten materiellen punkten.
\newblock {\em Wiener Berichte}, 58, (W.A.,\#5):517--560, (49--96), 1868.

\bibitem{Bo868-b}
L.~Boltzmann.
\newblock L{\"o}sung eines mechanischen problems.
\newblock {\em Wiener Berichte}, 58, (W.A.,\#6):1035--1044, (97--105), 1868.

\bibitem{Fe923}
E.~Fermi.
\newblock Dimostrazione che in generale un sistema meccanico \`e quasi
  ergodico.
\newblock {\em Nuovo Cimento}, 25:267--269, 1923.

\bibitem{Ko954}
A.N. Kolmogorov.
\newblock On the preservation of conditionally periodic motions.
\newblock {\em In Lecture Notes in Physics, Stochastic behavior in classical
  and quantum Hamiltonians, ed. G. Casati, J. Ford, Vol. 93, 1979}, 93, 1979.

\bibitem{Fo992}
J.~Ford.
\newblock The fermi-pasta-ulam problem: paradox turns discovery.
\newblock {\em Physics Reports}, 213:271--310, 1992.

\bibitem{Bo866}
L.~Boltzmann.
\newblock {\"U}ber die mechanische {B}edeutung des zweiten {H}auptsatzes der
  {W\"a}rme\-theorie.
\newblock {\em Wiener Berichte}, 53, (W.A.,\#2):195--220, (9--33), 1866.

\bibitem{Da014}
O.~Darrigol.
\newblock {\em {Physics \& Necessity}}.
\newblock Oxford University Press, Oxford, 2014.

\bibitem{Ma879}
J.~C. Maxwell.
\newblock On {B}oltzmann's theorem on the average distribution of energy in a
  system of material points.
\newblock {\em Transactions of the Cambridge Philosophical Society},
  12:547--575, 1879.

\bibitem{Cl871}
R.~Clausius.
\newblock {Ueber die Zur{\"u}ckf{\"u}hrung des zweites Hauptsatzes der
  mechanischen W{\"a}rmetheorie und allgemeine mechanische Prinzipien}.
\newblock {\em Annalen der Physik}, 142:433--461, 1871.

\bibitem{Ga025}
G.~Gallavotti.
\newblock {\em Nonequilibrium and irreversibility (II edition)}, volume LNP
  1040 of {\em Lecture Notes in Physics}.
\newblock Springer-Verlag, 2025.

\bibitem{Gi902}
J.~Gibbs.
\newblock {\em Elementary principles in statistical mechanics}.
\newblock Schribner, Cambridge, 1902.

\bibitem{Bo871-b}
L.~Boltzmann.
\newblock Einige allgemeine s{\"a}tze {\"u}ber {W\"a}rme\-gleichgewicht.
\newblock {\em Wiener Berichte}, 63, (W.A.,\#19):679--711, (259--287), 1871.

\bibitem{Ga016}
G.~Gallavotti.
\newblock Ergodicity: a historical perspective. equilibrium and nonequilibrium.
\newblock {\em European Physics Journal H}, 41,:181--259, 2016.

\bibitem{GJ020}
G.~Gallavotti and I.~Jauslin.
\newblock {A Theorem on Ellipses, an Integrable System and a Theorem of
  Boltzmann.}
\newblock {\em arXiv}, 2008.01955:1--9, 2020.

\bibitem{Fe021}
G.~Felder.
\newblock {Poncelet Property and Quasi-periodicity of the Integrable Boltzmann
  System}.
\newblock {\em Letters in Mathematical Physics}, 111:1--19, 2021.

\bibitem{Ga004c}
G.~Gallavotti.
\newblock {\em Classical mechanics and the quantum revolution in Fermi’s
  early works}, volume {In {\it Enrico Fermi: his work and legacy}, editors C.
  Bernardini and L. Bonolis}.
\newblock Springer, 2004.

\bibitem{Ni971}
T.~Nishida.
\newblock A note on an existence of conditionally periodic oscillation in a
  one-dimensional anharmonic lattice.
\newblock {\em Mem. Fac. Engrg. Kyoto Univ.}, 33:27–34, 1971.

\bibitem{Ri006}
B.~Rink.
\newblock Proof of {N}ishida's conjecture on anharmonic lattices.
\newblock {\em Communications Mathematical Physics}, 261:613--627, 2006.

\bibitem{Ga986}
G.~Gallavotti.
\newblock Quasi integrable mechanical systems.
\newblock {\em Phenom\`enes Critiques, Syst\`emes aleatories, Th\'eories de
  jauge, Proceedings, Les Houches, XLIII (1984), North Holland, Amsterdam},
  II:539--624, 1986.

\bibitem{Ar963b}
V.~Arnold.
\newblock {Small denominators and problems of stability of motion in classical
  and celestial mechanics}.
\newblock {\em Russian Mathematical Surveys}, 18:85--191, 1963.

\bibitem{Mo962}
J.~Moser.
\newblock On invariant curves of an area preserving mapping of the annulus.
\newblock {\em {Nachrichten Akademie Wissenshaften G{\"o}ttingen, II}},
  11:1--20, 1962.

\bibitem{BGGS984}
G.~Benettin, L.~Galgani, A.~Giorgilli, and J.~Strelcyn.
\newblock {A proof of Kolmogorov's theorem on invariant tori using canonical
  transformations defined by the Lie method}.
\newblock {\em Nuovo Cimento B}, 79:201--223, 1984.

\bibitem{Ga019b}
G.~Gallavotti.
\newblock Quasi periodic hamiltonian motions, scale invariance, harmonic
  oscillators.
\newblock {\em Journal of Mathematical Physics}, 60:062901, 2019.

\bibitem{Ga994b}
G.~Gallavotti.
\newblock {Twistless KAM tori}.
\newblock {\em Communications in Mathematical Physics}, 164:145--156, 1994.

\bibitem{Ga008e}
G.~Gallavotti.
\newblock {\em {The Fermi-Pasta-Ulam problem. A status report}}.
\newblock {Lecture Notes in Physics, ed. G. Gallavotti}. Springer, 2008.

\bibitem{Ri001}
B.~Rink.
\newblock Symmetry and resonance in periodic fpu chains.
\newblock {\em Communications Mathematical Physics}, 218:665–685, 2001.

\bibitem{Ga993}
G.~Gallavotti.
\newblock Some rigorous results about {3D} {N}avier {S}tokes.
\newblock {\em Les Houches 1992 NATO-ASI meeting on Turbulence in spatially
  extended systems, ed. by {R. Benzi, C. Basdevant, S. Ciliberto, Nova Science
  Publishers, NY}}, pages 45--81, 1993.

\bibitem{Ri008}
B.~Rink.
\newblock {\em An Integrable Approximation for the Fermi–Pasta–Ulam
  Lattice}, volume 728 of {\em Lecture Notes in Physics}.
\newblock Springer, 2008.

\bibitem{Fl974}
H.~Flaschka.
\newblock The toda lattice. ii. existence of integrals.
\newblock {\em Physical Review B}, 9:1924–1925, 1974.

\bibitem{He974}
M.~Henon.
\newblock Integrals of the toda lattice.
\newblock {\em Physical Review B}, 9:1921–1923, 1974.

\bibitem{HK006}
A.~Henrici and T.~Kappeler.
\newblock Birkhoff normal form for the periodic toda lattice.
\newblock {\em arXiv preprint nlin/0609045}, 2006.

\bibitem{Mo975b}
J.~Moser.
\newblock Finitely many mass points on the line under the influence of an
  exponential potential—an integrable system.
\newblock {\em Dynamical Systems, Theory and Applications: Battelle Seattle
  1974 Rencontres}, 38:467–497, 1975.

\bibitem{Ga008}
G.~Gallavotti.
\newblock {\em The Elements of Mechanics (II edition)}.
\newblock Gallavotti website [I edition was Springer 1984], Roma, 2008.

\end{thebibliography}

\end{document}